\documentclass[journal,twocolumn]{IEEEtran}
\usepackage[cmex10]{amsmath}
\usepackage{amssymb}
\usepackage{cite}
\usepackage{graphicx}
\usepackage{array,color}
\usepackage{multirow}
\usepackage{amsmath}
\usepackage{stfloats}
\usepackage{graphicx}
\usepackage{subfigure}
\usepackage{tabularx}
\usepackage{epsfig,epsf,color,balance,cite}
\usepackage{setspace}
\usepackage{bm}
\usepackage{textcomp}
\usepackage{algorithmic}
\usepackage{algorithm}
\usepackage{caption} 

\usepackage{hyperref}
\usepackage{comment}
\usepackage{subfigure}

\usepackage{caption}
\usepackage{graphicx}
\usepackage{setspace}
\usepackage{diagbox}

\usepackage{stfloats}
\fnbelowfloat
\usepackage{multirow}
\usepackage{graphicx}
\makeatletter

\newcommand{\Rmnum}[1]{\expandafter\@slowromancap\romannumeral #1@}
\makeatother
\usepackage{booktabs}
\usepackage{amsmath}

\makeatother
\pdfoutput=1
\begin{document}

\title{Large-Model AI for Near-Field Beam Prediction: A CNN–GPT2 Framework for 6G XL-MIMO}
\author{
Wang Liu, Cunhua Pan, $\textit{Senior Member, IEEE}$, Hong Ren, $\textit{Member, IEEE}$, Wei Zhang, $\textit{Fellow, IEEE}$, Cheng-Xiang Wang, $\textit{Fellow, IEEE}$, and Jiangzhou Wang, $\textit{Fellow, IEEE}$ \thanks{
	\emph{Corresponding author: Cunhua Pan.}
	

Wang Liu, Cunhua Pan, Hong Ren and Jiangzhou Wang are with National Mobile Communications Research Laboratory, Southeast University, Nanjing 211189, China. (e-mail: {w\_liu, cpan, hren, j.z.wang}@seu.edu.cn).

Wei Zhang is with the School of Electrical Engineering and Telecommunications, the University of New South Wales, Sydney NSW 2052, Australia (e-mail:
w.zhang@unsw.edu.au).

Cheng-Xiang Wang is with the National Mobile Communications Research Laboratory, School of Information Science and Engineering, Southeast University, Nanjing 211189, China, and also with the Pervasive Communication Research Center, Purple Mountain Laboratories, Nanjing 211111, China (e-mail: chxwang@seu.edu.cn).

 } }

\maketitle

\begin{abstract}
	The emergence of extremely large-scale antenna arrays (ELAA) in millimeter-wave (mmWave) communications, particularly in high-mobility scenarios, highlights the importance of near-field beam prediction. Unlike the conventional far-field assumption, near-field beam prediction requires codebooks that jointly sample the angular and distance domains, which leads to a dramatic increase in pilot overhead. Moreover, unlike the far-field case where the optimal beam evolution is temporally smooth, the optimal near-field beam index exhibits abrupt and nonlinear dynamics due to its joint dependence on user angle and distance, posing significant challenges for temporal modeling. To address these challenges, we propose a novel \textbf{Convolutional Neural Network–Generative Pre-trained Transformer 2 (CNN–GPT2) based near-field beam prediction framework}. Specifically, an uplink pilot transmission strategy is designed to enable efficient channel probing through widebeam analog precoding and frequency-varying digital precoding. The received pilot signals are preprocessed and passed through a CNN-based feature extractor, followed by a GPT-2 model that captures temporal dependencies across multiple frames and directly predicts the near-field beam index in an end-to-end manner. A pretraining--finetuning strategy is further adopted, where the model is first pretrained via masked prediction and then finetuned for the downstream beam prediction task, significantly improving training efficiency and accuracy. Simulation results under the 3GPP TR~38.901 channel model demonstrate that the proposed method achieves higher beam prediction accuracy than conventional recurrent models, while maintaining competitive performance in normalized beamforming gain. These results confirm the feasibility of employing large-model AI for robust and low-overhead near-field beam management in future 6G systems.
\end{abstract}

\begin{IEEEkeywords}
ELAA, near field, beam prediction, mmWave, CNN, GPT2.
\end{IEEEkeywords}
\vspace{-0.4cm}

\section{Introduction}

Millimetre-wave (mmWave) bands provide abundant spectrum resources and enable ultra-high data rates, making them a cornerstone of current fifth-generation (5G) and future sixth-generation (6G) mobile communication systems \cite{mmwave_survey, mmwave_1,cxwang1,cxwang2}. However, mmWave signals suffer from severe propagation loss compared to sub-6~GHz bands. To overcome this challenge, 5G systems deploy massive multiple-input multiple-output (MIMO) antenna arrays at the base station (BS) to establish highly directional transmissions \cite{bm_surey}
Looking ahead to 6G, the scale of antenna deployment is expected to further increase, leading to the concept of extremely large-scale antenna arrays (ELAA) \cite{xlmimo}. With a significantly larger number of antenna elements, ELAA can achieve sharper spatial focusing and higher array gains, thereby offering unprecedented improvements in spectral efficiency \cite{near_field_survey1,cui_1}.  

To fully exploit the potential of ELAA systems, accurate beam alignment between the BS and users is crucial\cite{bm_surey}. Conventional codebook-based beam training enables optimal beam selection but suffers from high-mobility scenarios, where rapid user movement causes trained beams to become quickly outdated, resulting in severe misalignment and throughput loss. To overcome this limitation, beam prediction has been introduced, leveraging temporal correlations in historical pilot signals to infer future optimal beams with reduced delay\cite{bp_survey}.


However, beam prediction in ELAA systems presents several key challenges. Firstly, the transition from the far-field to the near-field regime fundamentally alters the array response: while far-field beamforming depends solely on the angular domain under planar-wave assumptions, near-field propagation involves spherical wavefronts, making the array response jointly dependent on user angle and distance. As a result, the near-field beamforming codebook expands from a one-dimensional angular space to a two-dimensional angle–distance domain, dramatically increasing the number of codewords and the required pilot overhead. This enlarged codebook substantially raises pilot overhead and hinders fast beam adaptation under user mobility\cite{cui_1,cui_2}.


Secondly, from the perspective of temporal modeling, beam prediction in ELAA systems presents unique challenges. In far-field systems, the optimal beam index primarily depends on the user’s angular. As the user moves, the angle of arrival or departure changes gradually, leading to a smooth evolution of the optimal beam index over time. In contrast, in the near-field regime, the optimal beam index depends jointly on both the angle and the distance of the user. This dual dependence introduces strong spatial nonlinearity into the beam dynamics. Even small user movements can cause significant phase shifts across the antenna array, resulting in abrupt changes in the optimal near-field beam index. Consequently, the temporal evolution of the beam becomes highly non-smooth and less predictable, making it substantially more difficult for conventional time-series models to capture long-term dependencies\cite{near_field_survey2}.

\vspace{-0.2cm}
\subsection{Related Work}
\vspace{-0.2cm}

In recent years, various advanced signal processing and learning-based techniques have been investigated to enhance beam prediction in mmWave communication systems\cite{kf1,kf2,uav_beam1,uav_beam2,ai_for_bp,make_pre1,make_pre2,make_pre3,hand_blockage,integrate,attention_lstm,llm_bp}. For signal processing–based methods, extended kalman filter (EKF)-assisted beam tracking schemes have been proposed in \cite{kf1,kf2}, which combine spatial information, channel estimation, and minimum mean squared error (MMSE)  beamforming to enhance tracking accuracy and maintain reliable alignment under mobility.
Gaussian process regression-based beam prediction algorithms have been developed for unmanned aerial vehicle (UAV) communications \cite{uav_beam1,uav_beam2}. These methods enable adaptive beam design and reconstruction, thereby alleviating beam misalignment problems under high-mobility conditions.

Morevoer, artificial intelligence (AI) and machine learning (ML) have recently been introduced to further improve beam prediction efficiency and robustness. \cite{ai_for_bp} proposed practical AI/ML-based time-domain beam prediction schemes aligned with 3GPP Release-18, incorporating user mobility and validated through 3GPP-compliant over-the-air (OTA) tests, demonstrating the real-world feasibility of AI-assisted beam prediction.


Meanwhile, a deep learning framework that fuses sub-6 GHz channel state information (CSI) with low-overhead mmWave measurements was proposed in \cite{make_pre1}, achieving higher beamforming gain with reduced pilot overhead. 
Similarly, a multi-stage network (MSN) combining convolutional neural network (CNN) and Transformer modules with adaptive classifier selection was presented in \cite{make_pre2}, providing high accuracy while maintaining low computational complexity. 
In \cite{make_pre3}, three deep learning–assisted wide-beam training schemes were introduced, leveraging CNN and long short-term memory (LSTM) architectures to predict and calibrate narrow beams with improved robustness and reduced training overhead. 
In addition to these developments, a shift-Transformer model for user-side beam prediction under hand blockage was developed in \cite{hand_blockage}, where attention operations were replaced by shift mechanisms to lower complexity without sacrificing prediction accuracy. 
The authors of \cite{integrate} further proposed a compact neural network for joint probing-beam selection and beam prediction, integrating a differentiable sampling network with CNN and self-attention modules to achieve high prediction accuracy while minimizing model parameters. 
Furthermore, an attention-enhanced LSTM architecture was presented in \cite{attention_lstm} to jointly predict beam indices and signal-to-noise ratio (SNR), effectively reducing signaling overhead and feedback latency in mmWave systems.
Conventional recurrent models (Recurrent Neural Network (RNN), LSTM, Gated Recurrent Unit (GRU)) can capture short-term temporal dependencies but struggle with vanishing gradients and limited receptive fields when modeling nonlinear, non-stationary beam dynamics. Since near-field beams exhibit abrupt variations due to coupled angle–distance dynamics, global temporal modeling is essential—making large model like GPT-2 a more suitable backbone than Conventional recurrent models.


More recently, large language model (LLM)-based frameworks have been introduced to reformulate beam selection as a time-series forecasting task \cite{llm_bp}. By incorporating cross-variable attention and prompt-as-prefix mechanisms, these methods show promising potential for knowledge transfer and context-aware beam prediction. However, they rely heavily on precise historical angle-of-arrival (AoA) information and overlook distance-dependent effects, which limits their applicability in near-field communication scenarios.

For the near-field domain, various approaches have been explored to alleviate the excessive pilot overhead \cite{luyu,youchangsheng_1,youchangsheng_2,shixu,near_beam1,my_letter}. In \cite{luyu}, a hierarchical beam training scheme tailored for near-field channels was developed, where wide-beam codewords were optimized for spatial focusing. To further reduce overhead, the two-phase beam training method in \cite{youchangsheng_1} first estimated the user angle using a discrete Fourier transform (DFT) codebook and then refined the range through a polar-domain codebook. Although efficient, its pilot overhead still scales with the number of antennas, making it less practical for extremely large-scale arrays. A related study \cite{youchangsheng_2} introduced a two-stage hierarchical search that first locates the approximate angle using a far-field codebook, followed by precise angle–range refinement with a near-field hierarchical codebook. In \cite{shixu}, the near-field channel was projected into spatial–angular and slope–intercept domains to enable hierarchical beam training, yet such schemes remain sensitive to noise and require considerable feedback. For terahertz (THz) systems, a near-field beam training strategy using a uniform circular array (UCA) was proposed in \cite{near_beam1}, exploiting its geometric characteristics to reduce codebook size and enhance beam coverage.

Recently, deep learning has also been introduced to near-field beam training. In \cite{my_letter,my_tcom1}, convolutional neural network (CNN) and graph neural network (GNN) were employed to predict beams directly from received signals, effectively lowering pilot overhead while maintaining accuracy. However, these methods focus solely on spatial-domain beam training and do not incorporate temporal modeling.


\vspace{-0.3cm}

\subsection{Main Contributions}

  \vspace{-0.1cm}
  
Motivated by the challenges of excessive pilot overhead, nonlinear beam dynamics, and insufficient temporal modeling capability in near-field extremely large-scale MIMO (XL-MIMO) systems, this paper proposes a large-model-driven end-to-end framework for near-field beam prediction. The proposed framework seamlessly integrates practical system modeling, efficient pilot design, large-model-based learning, and pre-training strategies to achieve efficient and robust beam management. The main contributions are summarized as follows:

\begin{itemize}
	\item \textbf{End-to-end near-field beam prediction framework:}
%
%
	We reformulate near-field beam training as a beam prediction problem and build an \emph{end-to-end trainable} network that directly maps historical pilot signals to the next near-field beam index. This approach eliminates the need for explicit channel estimation or intermediate parameters. By integrating feature extraction, temporal modeling, and beam decision into one coherent process, the framework avoides the error accumulation seen in multi-stage designs and achieves faster inference with better consistency.

\item \textbf{Practical system modeling and pilot design:} 

To ensure a realistic system evaluation, both the standardized 5G NR frame structure and the 3GPP TR 38.901 channel model are incorporated into our scheme. On this basis, we design an uplink pilot transmission scheme that jointly employs Zadoff–Chu (ZC) sequences, widebeam analog precoding, and frequency-dependent digital precoding. This design makes full use of spatial–frequency diversity, significantly reducing pilot overhead while preserving high prediction accuracy and maintaining computational efficiency.

\item \textbf{Large-model-driven spatiotemporal learning architecture:}  
	Considering the highly nonlinear and non-smooth temporal evolution of near-field beams, we adopt a hybrid CNN–GPT-2 architecture. The convolutional module extracts and compresses spatial information from high-dimensional pilot data, while  the GPT-2 module captures long-range temporal dependencies across multiple radio frames.  

\item \textbf{Pre-training and fine-tuning strategy for enhanced generalization:}  
We further introduce a Bidirectional Encoder Representations from Transformers (BERT)-inspired pre-training–fine-tuning paradigm. In pre-training, a temporal masking task with an 80/10/10 policy facilitates context-aware feature learning; during fine-tuning, the model specializes in next-frame beam prediction. This two-stage training enhances temporal modeling capability, leading to improved prediction accuracy and better generalization performance.
\end{itemize}


The remainder of this paper is organized as follows. Section~\ref{system_model} introduces the considered system model and formulates the beam prediction problem. Section~\ref{scheme} details the proposed CNN–GPT-2-based prediction framework, including its network architecture and processing pipeline. Section~\ref{simulation} presents and analyzes the simulation results, and Section~\ref{conclusion} concludes the paper with final remarks.

The main notations used in this paper are as follows. Bold lowercase and uppercase letters (e.g., $\mathbf{a}$ and $\mathbf{A}$) denote vectors and matrices, respectively; scalars and sets are represented by $a$ and $\mathcal{A}$. The $i$-th entry of $\mathbf{a}$ and the $(i,j)$-th entry of $\mathbf{A}$ are written as $[\mathbf{a}]_{i}$ and $[\mathbf{A}]_{i,j}$. The absolute value is denoted by $|\cdot|$, while $(\cdot)^{\ast}$, $(\cdot)^{\mathrm{T}}$, and $(\cdot)^{\mathrm{H}}$ represent the conjugate, transpose, and Hermitian transpose, respectively. A complex Gaussian variable with mean $\mu$ and variance $\sigma^{2}$ is expressed as $\mathcal{C}\mathcal{N}(\mu,\sigma^{2})$, and $|\cdot|_{\mathrm{F}}$ denotes the Frobenius norm.


\vspace{-0.2cm}

\section{System Model}\label{system_model}

\subsection{Downlink Signal Model}
Consider a time-division duplexing (TDD) based multi-user mmWave-orthogonal frequency division multiplexing (OFDM) communication system.
A BS equipped with a uniform linear array (ULA)
\footnote{
	The proposed framework can be naturally extended to uniform planar array (UPA) deployments. 
	In this case, the array response depends on azimuth, elevation and distance, and only a corresponding three-dimensional near-field codebook needs to be adopted. 
	Hence, the proposed CNN–GPT-2-based prediction model and pilot transmission strategy remain directly applicable, since the underlying spherical-wavefront propagation and spatial–frequency correlations are preserved.}
of $N_{\mathrm{BS}}$ antennas and $N_{\mathrm{RF}}$ radio-frequency (RF) chains simultaneously serves $U$ single-antenna users over $K$ orthogonal subcarriers. 
The condition $U \leq N_{\mathrm{RF}} \ll N_{\mathrm{BS}}$ is assumed to hold, and to reduce power consumption redundant RF chains are switched off, i.e., $N_{\mathrm{RF}}=U$ \cite{sun,hybrid_precoding_AA_1}.


Let $\mathbf{h}_{u,k}^{\mathrm{dl}}\in \mathbb{C}^{1\times N_{\mathrm{BS}}}$ denote the downlink (DL) channel from the BS to user $u$ on the $k$-th subcarrier, where $u=1,2,\ldots,U$ and $k=1,2,\ldots,K$. During DL transmission, the received signal of user $u$ on subcarrier $k$ is given by
\begin{equation}\label{dl_r_ofdm}
	\setlength\abovedisplayskip{3pt}
	\setlength\belowdisplayskip{3pt}
	\begin{aligned}
		r_{u,k}^{\mathrm{dl}} 
		&= \mathbf{h}_{u,k}^{\mathrm{dl}} \mathbf{F}_{\mathrm{RF}} \mathbf{F}_{\mathrm{BB},k} \mathbf{s}_{k} + n_{u,k}^{\mathrm{dl}} \\
		&= \mathbf{h}_{u,k}^{\mathrm{dl}} \mathbf{F}_{\mathrm{RF}} \sum\limits_{j=1}^{U} \mathbf{f}_{j,k}^{\mathrm{BB}} s_{j,k} + n_{u,k}^{\mathrm{dl}},
	\end{aligned}
\end{equation}
where $\mathbf{s}_{k}=\left[s_{1,k},s_{2,k},\ldots,s_{U,k}\right]^{\mathrm{T}}\in \mathbb{C}^{U\times 1}$ denotes the transmitted signal vector on subcarrier $k$, satisfying the power constraint $\mathbb{E}\!\left[\mathbf{s}_{k}\mathbf{s}_{k}^{\mathrm{H}}\right]=\frac{P_{\mathrm{dl}}}{UK}\mathbf{I}_{U}$. Here, $\mathbf{F}_{\mathrm{BB},k}=\left[\mathbf{f}_{1,k}^{\mathrm{BB}},\mathbf{f}_{2,k}^{\mathrm{BB}},\ldots,\mathbf{f}_{U,k}^{\mathrm{BB}}\right]\in \mathbb{C}^{U\times U}$ is the digital precoder on subcarrier $k$, while $\mathbf{F}_{\mathrm{RF}}=\left[\mathbf{f}_{1}^{\mathrm{RF}},\mathbf{f}_{2}^{\mathrm{RF}},\ldots,\mathbf{f}_{U}^{\mathrm{RF}}\right]\in \mathbb{C}^{N_{\mathrm{BS}}\times U}$ is the frequency-flat analog precoder shared across all subcarriers. Moreover, $n_{u,k}^{\mathrm{dl}}\sim \mathcal{CN}(0,\sigma_{\mathrm{dl}}^{2})$ represents the additive white Gaussian noise (AWGN) at user $u$ on subcarrier $k$.

Since $\mathbf{F}_{\mathrm{RF}}$ is implemented via phase shifters, each of its entries has a constant modulus normalized as $\big|\left[\mathbf{F}_{\mathrm{RF}}\right]_{m,n}\big|^{2}=\frac{1}{N_{\mathrm{BS}}}$. Specifically, the elements take the form $\left[\mathbf{F}_{\mathrm{RF}}\right]_{m,n}=\tfrac{1}{\sqrt{N_{\mathrm{BS}}}}e^{j\phi_{m,n}}$, where $\phi_{m,n}$ is a quantized phase \cite{sun,hybrid_precoding_AA_1}. Furthermore, the digital precoders are required to satisfy the normalization constraint
\begin{equation}\label{bb_norm}
	\left\|\mathbf{F}_{\mathrm{RF}}\mathbf{f}_{u,k}^{\mathrm{BB}}\right\|_{\mathrm{F}}^{2}=1, \quad \forall u=1,2,\ldots,U,~k=1,2,\ldots,K,
\end{equation}
such that no additional power gain is introduced by the digital domain.


\vspace{-0.6cm}
\subsection{Channel Model}

In this work, the wireless propagation channel is modeled following the 3GPP TR~38.901 specification \cite{TR38901}, which captures the spatial and temporal characteristics of mmWave propagation through a geometry-based stochastic approach. The channel between the $s$-th transmit antenna element at the BS and the $n$-th receive antenna element at the user over delay $\tau$ and time $t$ can be expressed as a combination of the line-of-sight (LoS) and non-line-of-sight (NLoS) components, given by
\begin{equation}\label{tr_channel_1}
	\begin{aligned}
H_{n, s}(\tau, t)&=\sqrt{\frac{1}{K_R+1}}\,H_{n, s}^{\mathrm{NLOS}}(\tau, t)+\\
&\sqrt{\frac{K_R}{K_R+1}}\,H_{n, s, 1}^{\mathrm{LOS}}(t)\,\delta\!\left(\tau-\tau_1\right),
		\end{aligned}
\end{equation}
where $K_R$ is the Rician factor, and $\tau_1$ denotes the delay of the LoS path. 

\addtocounter{equation}{1}
The LoS component $H_{n, s, 1}^{\mathrm{LOS}}(t)$ is formulated as (\ref{los}), which is shown at the top of the next page.
\begin{figure*}[!t]
	\vspace{-0.3cm}
	\hrulefill
	\vspace{-0.1cm}
	\setcounter{equation}{3}
	\begin{align}\label{los}
			H_{n, s, 1}^{\mathrm{LOS}}(t)
		&=\left[
		\begin{array}{c}
			F_{rx,n,\theta}(\theta_{\mathrm{LOS,ZOA}},\phi_{\mathrm{LOS,AOA}})\\
			F_{rx,n,\phi}(\theta_{\mathrm{LOS,ZOA}},\phi_{\mathrm{LOS,AOA}})
		\end{array}
		\right]^{\!\mathrm{T}}
		\!\!\!
		\begin{bmatrix}
			1 & 0 \\[1pt]
			0 & -1
		\end{bmatrix}
		\left[
		\begin{array}{c}
			F_{tx,s,\theta}(\theta_{\mathrm{LOS,ZOD}},\phi_{\mathrm{LOS,AOD}})\\
			F_{tx,s,\phi}(\theta_{\mathrm{LOS,ZOD}},\phi_{\mathrm{LOS,AOD}})
		\end{array}
		\right] \nonumber\\
		&\quad\cdot 
		\exp\!\left(-j\frac{2\pi d_{3\mathrm{D}}}{\lambda_0}\right)
		\exp\!\left(j\frac{2\pi \hat{r}_{rx,\mathrm{LOS}}^{\mathrm{T}}\!\cdot\!\bar{d}_{rx,n}}{\lambda_0}\right)
		\exp\!\left(j\frac{2\pi \hat{r}_{tx,\mathrm{LOS}}^{\mathrm{T}}\!\cdot\!\bar{d}_{tx,s}}{\lambda_0}\right)
		\exp\!\left(j\frac{2\pi \hat{r}_{rx,\mathrm{LOS}}^{\mathrm{T}}\!\cdot\!\bar{v}}{\lambda_0}t\right),
	\end{align}
\vspace{-0.8cm}
	\setcounter{equation}{4}
\end{figure*}

In (\ref{los}), $F_{rx}$ and $F_{tx}$ represent the antenna field patterns at the receiver and transmitter, respectively; $\theta$ and $\phi$ denote the zenith and azimuth angle, respectively; $d_{3\mathrm{D}}$ is the three-dimensional distance between the transmitter and receiver; $\hat{r}_{rx,\mathrm{LOS}}$ and $\hat{r}_{tx,\mathrm{LOS}}$ are the unit direction vectors of arrival and departure; $\bar{d}_{rx,n}$ and $\bar{d}_{tx,s}$ denote the antenna element positions at the user and BS sides, and $\bar{v}$ denotes the user velocity vector. 

For the NLoS component $H_{n,s}^{\mathrm{NLOS}}(\tau,t)$, multiple scattering clusters and sub-paths are considered according to TR~38.901, while its full expression is omitted here due to space limitations. The detailed mathematical formulation can be found in the official specification \cite{TR38901}.

\subsubsection{OFDM-domain channel representation}
The time–delay channel model in \eqref{tr_channel_1} can be transformed to the OFDM subcarrier domain by taking the Fourier transform with respect to delay. Let the subcarrier spacing be $\Delta f$ and the $k$-th subcarrier frequency be $f_k = f_c + k\Delta f$ with $k=0,1,\dots,K-1$. The frequency-domain channel between the $s$-th transmit element and the $n$-th receive element on subcarrier $k$ at time $t$ is given by
\begin{equation}\label{freq_domain_cont}
	H_{n,s}[k,t]
	= \int_{-\infty}^{\infty} H_{n,s}(\tau,t)\, e^{-j2\pi f_k \tau}\, \mathrm{d}\tau .
\end{equation}
Substituting \eqref{tr_channel_1} into \eqref{freq_domain_cont} and using the fact that the LoS term contains a Dirac impulse at $\tau=\tau_1$ yields
\begin{equation}\label{freq_domain_split}
	\begin{aligned}
		H_{n,s}[k,t]
		&= \sqrt{\frac{1}{K_R+1}}\; \underbrace{\int H_{n,s}^{\mathrm{NLOS}}(\tau,t)\,e^{-j2\pi f_k \tau}\,\mathrm{d}\tau}_{\displaystyle H_{n,s}^{\mathrm{NLOS}}[k,t]} \\
		&\quad + \sqrt{\frac{K_R}{K_R+1}}\; H_{n,s,1}^{\mathrm{LOS}}(t)\, e^{-j2\pi f_k \tau_1}.
	\end{aligned}
\end{equation}

In practical channel generation and simulation, the NLoS component is commonly represented by a finite set of discrete delay taps (clusters/subpaths). Let the NLoS part be modeled as
\[
H_{n,s}^{\mathrm{NLOS}}(\tau,t)=\sum_{p=1}^{P} \beta_{p}(t)\,\delta(\tau-\tau_p),
\]
where $\beta_p(t)$ and $\tau_p$ denote the complex gain and the delay of the $p$-th NLoS tap (cluster or subpath). The OFDM-frequency-domain NLoS contribution then becomes
\begin{equation}\label{nlos_sum}
	H_{n,s}^{\mathrm{NLOS}}[k,t]
	= \sum_{p=1}^{P} \beta_{p}(t)\, e^{-j2\pi f_k \tau_p}.
\end{equation}
Combining \eqref{freq_domain_split} with \eqref{nlos_sum}, the per-subcarrier channel used in the OFDM system is
\begin{equation}\label{final_ofdm_channel}
	\begin{aligned}
	H_{n,s}[k,t]
	&= \sqrt{\frac{1}{K_R+1}}\sum_{p=1}^{P} \beta_{p}(t)\, e^{-j2\pi f_k \tau_p}\\
	&+ \sqrt{\frac{K_R}{K_R+1}}\, H_{n,s,1}^{\mathrm{LOS}}(t)\, e^{-j2\pi f_k \tau_1}.
\end{aligned}
\end{equation}


Since each user is equipped with a single antenna, the receive-antenna index $n$ can be dropped. Denote by $H_{s}[k,t]\triangleq H_{n_0,s}[k,t]$ the frequency-domain channel between the $s$-th BS antenna element and the (single-antenna) user on subcarrier $k$ at time $t$, where $n_0$ indicates the only receive element. The BS-to-user channel thus admits the vector form
\begin{equation}\label{vec_channel}
	\mathbf{h}[k,t]=\big[H_{1}[k,t],\,H_{2}[k,t],\,\ldots,\,H_{N_{\mathrm{BS}}}[k,t]\big]\in\mathbb{C}^{1\times N_{\mathrm{BS}}},
\end{equation}
which is the frequency-domain row-vector channel used in the OFDM-based downlink expressions (e.g., \eqref{dl_r_ofdm}) by replacing $\mathbf{h}_{u,k}^{\mathrm{dl}}$ with $\mathbf{h}[k,t]$ when referring to the considered user.

\subsection{Problem Formulation}
 The key task is to determine the analog precoder $\mathbf{F}_{\mathrm{RF}}$ from a near-field codebook, while the digital precoder can be obtained using conventional methods such as  zero-forcing (ZF) or minimum mean-squared error (MMSE). Due to hardware constraints, each column of $\mathbf{F}_{\mathrm{RF}}$ must be selected from a finite codebook. Unlike the discrete Fourier transform (DFT)-based far-field codebook, the near-field codebook is constructed under the spherical-wave assumption, incorporating both angular and distance domains \cite{near_field_survey1,near_field_survey2,cui_1,cui_2}. Specifically, the codebook is defined as
\begin{equation}\label{near_codebook}
	\mathcal{N} = \big\{ \mathbf{b}(\psi_{s},r_{m}) ~\big|~ s=1,\ldots,N_{\mathrm{BS}};\ m=1,\ldots,M \big\},
\end{equation}
where $\psi_s$ and $r_m$ denote the sampled angle and distance, respectively. The steering vector $\mathbf{b}(\psi_s,r_m)\in\mathbb{C}^{N_{\mathrm{BS}}\times 1}$ associated with the direction $\psi_s$ and distance $r_m$ is expressed as
\begin{equation}\label{steer_vector}
	\mathbf{b}(\psi_s,r_m) = \frac{1}{\sqrt{N_{\mathrm{BS}}}}
	\left[
	e^{j\frac{2\pi}{\lambda}\!\left(r_m^{(1)}-r_m\right)},
	\ldots,
	e^{j\frac{2\pi}{\lambda}\!\left(r_m^{(N_{\mathrm{BS}})}-r_m\right)}
	\right]^{T},
\end{equation}
where $r_m^{(s)}$ denotes the distance from the $s$-th antenna element ($s=1,2,\ldots,N_{\mathrm{BS}}$) to the sampling point $(\psi_s,r_m)$, and $\lambda$ is the carrier wavelength.

The dowmlink analog precoder design can thus be formulated as a beam training problem. For user $u$, the optimal codeword is chosen to maximize its effective beamforming gain across all $K$ subcarriers:
\begin{equation}\label{beamtraining}
	\underset{\mathbf{f}_{u}^{\mathrm{RF}}\in \mathcal{N}}{\max} \ 
	g_{u}, 
	\quad 
	g_{u}=\sum_{k=1}^{K}\big|\mathbf{h}_{u,k}^{\mathrm{dl}} \mathbf{f}_{u}^{\mathrm{RF}}\big|.
\end{equation}

However, the near-field codebook contains a significantly larger number of codewords compared to the far-field case, due to the additional distance dimension. Exhaustively searching for the optimal codeword not only incurs excessive pilot overhead but also lowers the likelihood of finding the best beam in practical systems. This challenge becomes even more critical for extremely large-scale antenna arrays (ELAA). To overcome this, we recast the beam selection process as a \textbf{end-to-end beam prediction problem}, where a large model is trained to directly predict the near-field codeword from the observed pilot signals, thereby reducing overhead while improving prediction accuracy.

\vspace{-0.2cm}

\section{GPT-2 based Near-field Beam Prediction Scheme}\label{scheme}

\subsection{5G NR Frame Structure and SRS Placement}

\begin{figure}[t]
	\centering
	\includegraphics[width=3.5in]{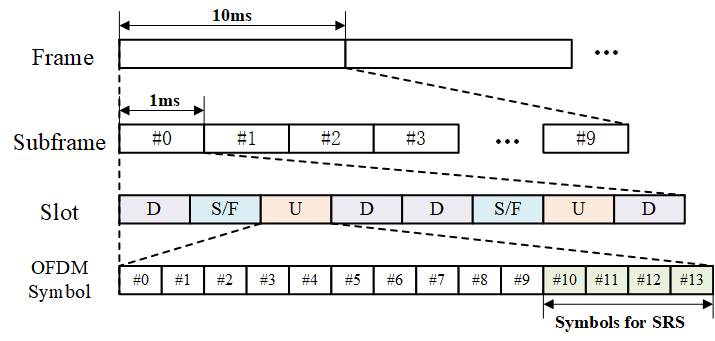}
	\vspace{-0.2cm}
	\caption{\fontsize{11pt}{\baselineskip}\selectfont Transmission organization in 5G NR.}
	\label{5gnr}
	\vspace{-0.6cm}
\end{figure}

To ensure that the proposed scheme is evaluated under realistic system assumptions, we adopt the standardized 5G New Radio (NR) frame structure specified by 3GPP as shown in Fig. \ref{5gnr}. In particular, a radio frame has duration of $10$\,ms and consists of $10$ subframes. With subcarrier spacing (SCS) of $120$\,kHz (numerology $\mu=3$), each subframe contains $2^{\mu}=8$ slots and each slot occupies $14$ OFDM symbols with normal cyclic prefix \cite{TS38211}. Hence, the slot duration is $0.125$\,ms and the frame contains $80$ slots in total \cite{TS38211,3GPP138211}.

In our design, the system operates in TDD mode with a periodic DL/UL pattern. In this work, we use a repeating \emph{D–S/F–U–D} arrangement over slots, where \emph{D}, \emph{U}, and \emph{S/F} denote downlink, uplink, and flexible (switching) slot types, respectively. The exact per-symbol DL/UL/Flexible allocation within each slot follows the 3GPP NR slot-format framework \cite{TS38211}. This setting captures the symbol-level flexibility of NR while matching practical deployments.

For uplink channel sounding, we transmit the sounding reference signal (SRS) in the last $n$ OFDM symbols of each \emph{U} slot, enabling the BS to estimate wideband CSI and perform near-field beam prediction. According to 3GPP TS~38.211, the SRS time-domain allocation resides within the last six symbols of a slot; common configurations allocate $N_\text{symb}^\text{SRS}\in\{1,2,4\}$ symbols starting from symbol index $l_0\in\{8,\ldots,13\}$ so that the SRS remains in the slot tail \cite{TS38211}. In this paper, we set
\[
n = N_\text{symb}^\text{SRS} = 4,
\]
i.e., SRS occupies the last four OFDM symbols of each uplink slot, which is standard-compliant and widely used in practice \cite{3GPP13810101}.%

The above frame and SRS design ensures that our near-field beam prediction uses measurements consistent with NR physical-layer timing and resource mapping while keeping the pilot overhead low. Under this configuration, a radio frame contains $80$ slots and thus provides a total of $80\times8=640$ OFDM symbols available for SRS transmission per frame.

\subsection{Uplink Pilot Transmission Strategy}
To enable the BS to acquire sufficient channel information for near-field beam prediction, we design a practical uplink pilot transmission strategy that includes the pilot sequence design, the analog precoder design, and the digital precoder design. The overall idea is to exploit the spatial and frequency diversity of the channel across subcarriers and RF chains, so that the BS can efficiently measure the channel response corresponding to a wide set of probing beams.
It is worth noting that the strategy described here corresponds to the pilot transmission scheme within a single radio frame.

\subsubsection{Pilot Sequence Design}
Each user transmits sounding reference signals (SRS) during the last four OFDM symbols of the uplink slot. To ensure mutual orthogonality among users, different SRS ports are employed so that pilots are multiplexed in the code domain. For each user, the pilot sequence on each OFDM symbol is generated by a ZC sequence of length $K$, where $K$ is the number of subcarriers. The ZC root index is varied across OFDM symbols to further improve the orthogonality and frequency diversity. Specifically, the pilot matrix $\mathbf{X}^{\mathrm{zc}}\in \mathbb{C} ^{K\times T}$ pilot OFDM symbols is constructed as
\begin{equation}
	X_{k,t}^{\mathrm{zc}}=\frac{1}{\sqrt{K}}
	\exp\!\left(-j\pi r_t \frac{k(k+\eta)}{K}\right),
\end{equation}
where $r_t$ denotes the ZC root index for the $t$-th OFDM symbol, and $\eta=0$ (resp. $\eta=1$) when $K$ is odd (resp. even). The root indices $\{r_t\}$ are chosen to be coprime with $K$, ensuring good correlation properties.

\subsubsection{Analog Precoder Design}
During pilot transmission, the analog precoder is constructed from a widebeam codebook that is derived from the far-field DFT codebook. Specifically, each widebeam codeword is generated by combining $v$ adjacent DFT codewords and then normalizing the phase of each antenna element to satisfy the constant-modulus constraint imposed by hardware. The $g$-th widebeam codeword is given by
\begin{equation}
	\mathbf{f}_{\mathrm{RF},g}=\frac{1}{\sqrt{N_{\mathrm{BS}}}}
	\exp\!\Big(j\angle\!\Big(\frac{1}{v}\sum_{i=0}^{v-1}\mathbf{f}_{\mathrm{DFT},\, (g-1)v+i+1}\Big)\Big),
\end{equation}
where $\mathbf{f}_{\mathrm{DFT},m}$ denotes the $m$-th DFT codeword of length $N_{\mathrm{BS}}$, $v$ is the grouping factor, and $G$ is the total number of widebeam codewords. 

Since the BS is equipped with only $N_{\mathrm{RF}}$ RF chains, it can simultaneously activate at most $N_{\mathrm{RF}}$ widebeam codewords per OFDM symbol. Therefore, the widebeam codewords are grouped in order, and $N_{\mathrm{RF}}$ beams are probed in each OFDM symbol until the entire codebook is swept. The analog precoder corresponding to the $t$-th OFDM pilot symbol is then
\begin{equation}
	\mathbf{F}_{\mathrm{RF},t}=
	\big[\mathbf{f}_{\mathrm{RF},\,(t-1)N_{\mathrm{RF}}+1},\,
	\mathbf{f}_{\mathrm{RF},\,(t-1)N_{\mathrm{RF}}+2},\,
	\ldots,\,
	\mathbf{f}_{\mathrm{RF},\,tN_{\mathrm{RF}}}\big]
\end{equation}
where $t=1,\dots,\lceil G/N_{\mathrm{RF}}\rceil$. In this way, approximately $G/N_{\mathrm{RF}}$ OFDM symbols are required to exhaustively probe all widebeam codewords.


It is worth highlighting the motivation behind constructing the analog precoder from widebeam codewords. 
The received pilot signals under these widebeams contain sufficient spatial information to infer the near-field beam index, although their nonlinear mapping is highly complex\cite{make2, my_letter}. 
Compared with exhaustive narrow-beam scanning, the number of widebeam codewords $G$ is much smaller, which greatly reduces the overhead required to cover the entire angular space. 
 Moreover, these widebeams are designed to provide complete coverage of the spatial domain without the need to switch off subsets of antenna elements, which is often required in conventional widebeam codebooks. This property makes the proposed widebeam codebook more efficient and hardware-friendly for practical near-field channel sounding.

\subsubsection{Digital Precoder Design}
In addition to analog beam sweeping, we introduce frequency-varying digital precoding to further exploit the frequency-domain resources. The idea is to assign different orthogonal directions on different subcarriers, so that the BS receives richer channel observations.

Let $\mathbf{Q}\in\mathbb{C}^{N_{\mathrm{RF}}\times N_{\mathrm{RF}}}$ denote the a $N_{\mathrm{RF}}$-dimensional unitary DFT matrix, whose columns are $N_{\mathrm{RF}}$ mutually orthogonal basis vectors. For the $k$-th subcarrier and the $t$-th pilot OFDM symbol, the digital precoder is defined as
\begin{equation}
	\mathbf{F}_{\mathrm{BB},k,t} = \mathbf{Q}_{(:,\,\mathcal{I}_k)},
\end{equation}
where $\mathcal{I}_k$ is the index set of selected columns. To ensure that different subcarriers probe different combinations of directions, we use a cyclic-shift rule:
\begin{equation}
	\mathcal{I}_k = \Big\{\, (j+k-1 \bmod N_{\mathrm{RF}})+1 \;\big|\; j=0,1,\ldots,N_{\mathrm{RF}}-1 \,\Big\}.
\end{equation}

In other words, the $k$-th subcarrier takes a shifted version of the DFT basis compared with the $(k-1)$-th subcarrier. This guarantees that across the entire band, the system cycles through all orthogonal directions, thereby enhancing the diversity of pilot measurements without introducing extra signaling overhead.

\subsubsection{Received Pilot Signal}
By jointly applying the analog and digital precoders, the effective hybrid precoder at the BS is given by $\mathbf{F}_{\mathrm{hyb},k,t}=\mathbf{F}_{\mathrm{RF},t}\mathbf{F}_{\mathrm{BB},k,t}$. The received pilot signal at the BS corresponding to user $u$ on subcarrier $k$ and OFDM symbol $t$ can be expressed as
\begin{equation}
	\mathbf{y}_{u,k,t}
	= \sqrt{P_\text{ul}}\,
	\mathbf{F}_{\mathrm{hyb},k,t}^{H}\mathbf{h}_{u,k,t}^{\mathrm{ul}}\, X_{k,t}^{\mathrm{zc}}
	+ \mathbf{F}_{\mathrm{hyb},k}^{H}\mathbf{n}_{k,t},
\end{equation}
where $\mathbf{h}_{u,k}^{\mathrm{ul}}\in\mathbb{C}^{N_{\mathrm{BS}}\times 1}$ denotes the uplink channel of user $u$, $X_{k,t}^{\mathrm{zc}}$ is the transmitted pilot symbol, $P_\text{ul}$ represents the uplink pilot power, and $\mathbf{n}_{k,t}$ is the noise. 
Since a TDD system is considered, the uplink channel $\mathbf{h}_{u,k}^{\mathrm{ul}}$ is the transpose of the downlink channel $\mathbf{h}_{u,k}^{\mathrm{dl}}$ due to channel reciprocity. 
After sweeping over all widebeam codewords, the BS collects a measurement matrix $\mathbf{Y}_u\in\mathbb{C}^{K\times G}$ that captures the responses of user $u$ to all probing beams. 
This measurement matrix serves as the input feature for the subsequent large-model-based beam prediction scheme.



In summary, the proposed pilot scheme within one radio frame occupies $G/N_{\mathrm{RF}}$ OFDM symbols. The pilot measurement matrix obtained for the $p$-th radio frame is denoted by $\mathbf{Y}_{u,p}$, which will serve as the input feature for the subsequent large-model-based beam prediction scheme.

\vspace{-0.2cm}

\subsection{Proposed GPT-2 based Near-field Beam Prediction Scheme}
As discussed in (\ref{beamtraining}), the analog precoder design can be formulated as a beam training problem, where the goal is to select, for each user, the near-field codeword that maximizes its effective beamforming gain across all subcarriers. Unlike far-field cases with angular-only codebooks, near-field training requires joint angle–distance search, dramatically enlarging the codebook and causing prohibitive pilot overhead in ELAA systems.

To address this challenge, we turn to deep learning and large-model techniques, aiming to replace the costly search procedure with direct prediction from observed pilot signals. The motivation for applying large models, such as GPT-2, arises from their powerful sequence modeling capability. In particular, uplink pilot measurements across multiple frames exhibit strong temporal correlation, which can be effectively captured by autoregressive architectures. Furthermore, GPT-2 has demonstrated remarkable generalization ability in handling long sequential dependencies, making it a promising candidate for predicting future beamforming decisions. This motivates us to recast the beam training task as a sequence-to-label prediction problem, where the input is a sequence of pilot signals and the output is the index of the predicted near-field beam index.

\begin{figure*}[t]
	\centering
	\includegraphics[width=7.8in]{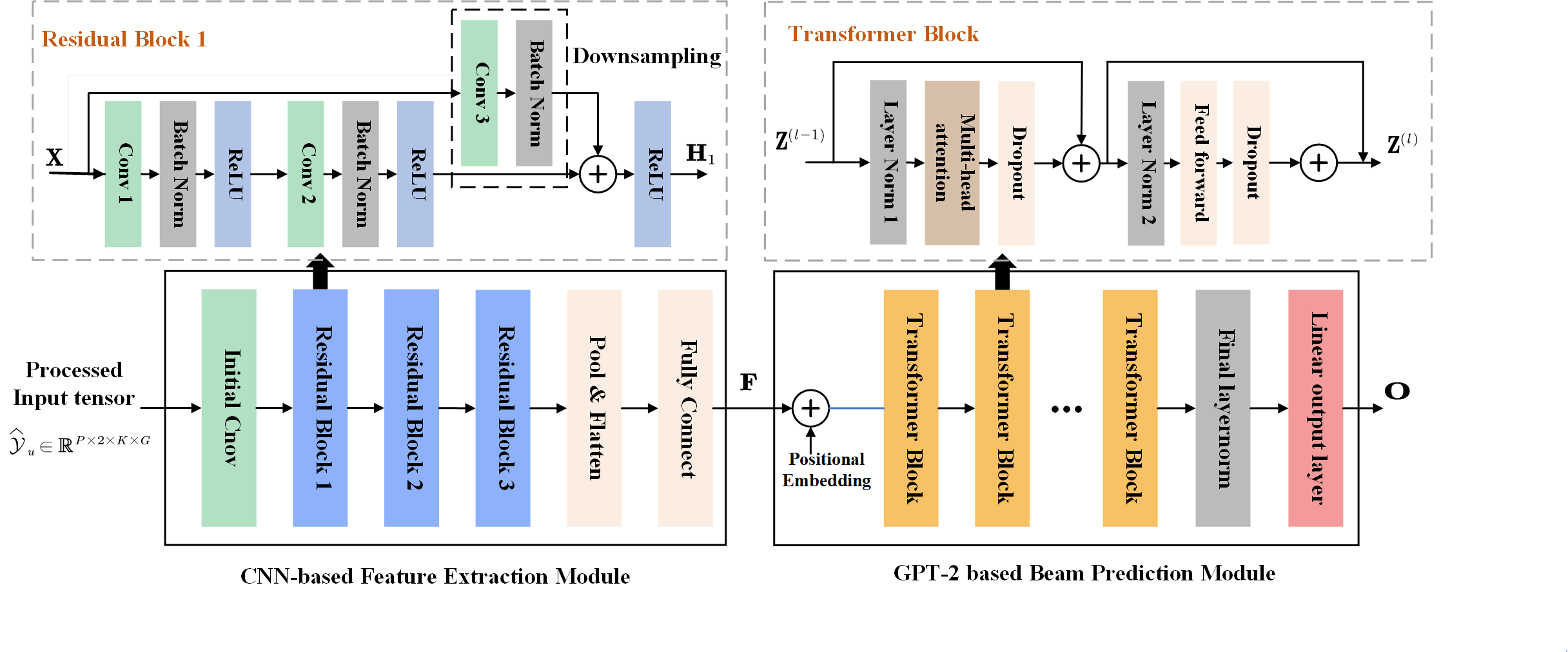}
	\vspace{-0.8cm}
	\caption{\fontsize{11pt}{\baselineskip}\selectfont An illustration of the nertwrok architecture of CNN-GPT-2.}
	\label{cnn_gpt2}
	\vspace{-0.6cm}
\end{figure*}

Specifically, we collect the uplink pilots from the past $P$ radio frames for each user, denoted as $\{\mathbf{Y}_{u,1},\mathbf{Y}_{u,2},\dots,\mathbf{Y}_{u,P}\}$, and feed them into a GPT-2 based prediction model. The model then directly outputs the index of the near-field codeword at the next radio frames. This process can be expressed mathematically as
\begin{equation}\label{beamtraining_gpt2}
	\left\{ \mathbf{b}_{u,P+1}^{\star} \right\} =f_{\mathrm{CNN-GPT-}2}\bigl( \mathbf{Y}_{u,1},\dots ,\mathbf{Y}_{u,P} \bigr) ,
\end{equation}
where $\mathbf{b}_{u,P+1}^{\star}$ denotes the predicted index of the predicted near-field beam index for user $u$ at the next radio frames, and $f_{\mathrm{CNN-GPT-2}}(\cdot)$ represents the mapping implemented by the CNN-GPT-2 model. The complete architecture of CNN-GPT-2 model is shown in Fig. \ref{cnn_gpt2}.

\subsubsection{Preprocessing Module}
Before feeding the pilot observations into the GPT-2 network, a preprocessing step is applied to ensure that the input is real-valued and properly normalized. Let the pilot sequence of user $u$ from the past $P$ radio frames be denoted as
\begin{equation}
	\mathcal{Y}_{u} = \big[\mathbf{Y}_{u,1}, \mathbf{Y}_{u,2}, \dots, \mathbf{Y}_{u,P}\big] 
	\in \mathbb{C}^{P \times K \times G}.
\end{equation}
Since $\mathcal{Y}_{u}$ is complex-valued, we first separate it into real and imaginary parts:
\begin{equation}
	\widetilde{\mathcal{Y}}_{u} = \big[\Re(\mathcal{Y}_{u}),\, \Im(\mathcal{Y}_{u})\big]
	\in \mathbb{R}^{P \times 2 \times K \times G}.
\end{equation}

To further enhance training stability and reduce the impact of amplitude variations, we apply per-sample normalization. Specifically, for each entry $z$ in $\widetilde{\mathcal{Y}}_{u}$, we compute
\begin{equation}
	\widehat{z} = \frac{z - \mu}{\sigma},
\end{equation}
where $\mu$ and $\sigma$ denote the mean and standard deviation of the corresponding sample, respectively. In this way, each processed input tensor has zero mean and unit variance, i.e.,
\begin{equation}
	\widehat{\mathcal{Y}}_{u} \in \mathbb{R}^{P \times 2 \times K \times G},
	\quad \mathbb{E}[\widehat{\mathcal{Y}}_{u}] = 0,\;\; 
	\mathrm{Var}[\widehat{\mathcal{Y}}_{u}] = 1.
\end{equation}

The normalized tensor $\widehat{\mathcal{Y}}_{u}$ is then passed to the subsequent CNN-based feature extraction and sequence modeling blocks of the GPT-2 based prediction framework.

\subsubsection{CNN-based Feature Extraction Module}
To effectively extract discriminative features from the preprocessed pilot tensors, we adopt a CNN with residual connections. The CNN is designed to compress the high-dimensional pilot input into a compact feature representation while preserving key spatial–frequency structures that are relevant for beam prediction. 

\paragraph{Residual Block} 
Each residual block consists of two convolutional layers with batch normalization (BN) and ReLU activation, followed by a skip connection to improve training stability. Given an input feature map $\mathbf{X}\in\mathbb{R}^{C\times H\times W}$, the residual block computes
\begin{equation}
	\mathrm{ResBlock}(\mathbf{X}) =\mathrm{ReLU}\!\big( \mathcal{F}(\mathbf{X};\Theta) + \mathcal{R}(\mathbf{X}) \big),
\end{equation}
where $\mathcal{F}(\cdot)$ denotes the stacked convolution–BN–ReLU layers with learnable parameters $\Theta$, and $\mathcal{R}(\cdot)$ is either the identity mapping or a $1\times 1$ convolution when downsampling is required. Dropout is further employed to enhance generalization.

\paragraph{CNN Feature Extractor.}
The overall CNN feature extractor first applies an initial convolution to expand the channel dimension (here, the term \emph{channel} refers to the input feature channels of the CNN, rather than the wireless communication channel), followed by three residual blocks:
\begin{align}
	\mathbf{H}_1 &= \mathrm{ResBlock}_1(\mathbf{X}), 
	\;\;\;\;\mathbf{H}_2 = \mathrm{ResBlock}_2(\mathbf{H}_1), \nonumber\\
	\mathbf{H}_3& = \mathrm{ResBlock}_3(\mathbf{H}_2).
\end{align}
Here $\mathrm{ResBlock}_1$ includes downsampling to reduce spatial dimensions by a factor of two, while $\mathrm{ResBlock}_2$ and $\mathrm{ResBlock}_3$ preserve the feature resolution.

\paragraph{Global Pooling and Projection}
To generate a fixed-size feature embedding, we apply adaptive average pooling to compress each feature map into a spatial grid of size $H_p \times W_p$:
\begin{equation}
	\mathbf{H}_{\mathrm{pool}} = \mathrm{Pool}(\mathbf{H}_3) \in \mathbb{R}^{C_p \times H_p \times W_p},
\end{equation}
where $C_p$ denotes the number of feature channels after the last residual block, and $H_p$, $W_p$ denote the height and width of the pooled feature maps, respectively.


The pooled features are flattened and passed through a fully connected layer with ReLU activation and dropout to obtain the final embedding of dimension $D_{\mathrm{out}}$:
\begin{equation}
	\mathbf{f} = \mathrm{ReLU}(\mathbf{W}\,\mathrm{vec}(\mathbf{H}_{\mathrm{pool}})+\mathbf{b}) 
	\in \mathbb{R}^{D_{\mathrm{out}}},
\end{equation}
where $\mathrm{vec}(\mathbf{H}_{\mathrm{pool}})$ denotes the vectorization of the pooled feature tensor 
$\mathbf{H}_{\mathrm{pool}}\in \mathbb{R}^{C_p \times H_p \times W_p}$ 
into a column vector of length $C_p H_p W_p$, 
$\mathbf{W}\in \mathbb{R}^{D_{\mathrm{out}} \times (C_p H_p W_p)}$ 
is the learnable weight matrix of the fully connected layer, 
and $\mathbf{b}\in \mathbb{R}^{D_{\mathrm{out}}}$ is the corresponding bias term.

\paragraph{Output Representation}
For $P$ input frames, the CNN operates on each $p$ sample individually and outputs a sequence of feature embeddings:
\begin{equation}
	\mathbf{F}=\left[ \mathbf{f}_1,\mathbf{f}_2,\dots \mathbf{f}_P \right] \in \mathbb{R} ^{P\times D_{\mathrm{out}}},
\end{equation}
which is then fed into the GPT-2 sequence modeling module. In this way, the CNN serves as a projection head that converts complex-valued pilot observations into compact, real-valued feature sequences suitable for temporal prediction.

\subsubsection{GPT-2 based Beam Prediction Module}
After feature extraction, the temporal modeling and codeword prediction are carried out by a GPT-2 style Transformer network. The GPT-2 model is particularly suitable for this task due to its strong capability of capturing long-range dependencies in sequential data, which in our case correspond to the temporal evolution of pilot measurements across multiple frames.

\paragraph{Input Embedding and Positional Encoding}
Let $\mathbf{F}\in \mathbb{R}^{ P\times D_{\mathrm{out}}}$ denote the feature sequence obtained from the CNN module.
Here, we set $D_{\mathrm{out}} = D_{\mathrm{emb}}$ so that the CNN output dimension matches the embedding dimension of the Transformer where $D_{\mathrm{emb}}$ is the Transformer embedding size, ensuring seamless integration between the two modules. 
For each time step $p$, a learnable positional embedding $\mathbf{e}_p\in \mathbb{R}^{D_{\mathrm{emb}}}$ is added to the input, yielding
\begin{equation}
	\mathbf{z}_p = \mathbf{f}_p + \mathbf{e}_p,
\end{equation}
where $\mathbf{f}_p$ is the CNN feature at step $p$. A dropout layer is then applied to improve generalization.

\paragraph{Transformer Blocks.}
The core of GPT-2 consists of stacked Transformer blocks. Each block is composed of a multi-head self-attention layer followed by a position-wise feed-forward network (FFN), with residual connections and layer normalization applied around both components. Denote the input to the $l$-th block as $\mathbf{Z}^{(l-1)}=\left[ \mathbf{z}_{1}^{\left( l-1 \right)},\dots ,\mathbf{z}_{P}^{\left( l-1 \right)} \right] ^{\mathrm{T}}\in \mathbb{R} ^{P\times D_{\mathrm{emb}}}$, then the block computes
\begin{equation}
	\mathbf{H}^{(l)} = \mathbf{Z}^{(l-1)} + 
	\mathrm{MHA}\!\big(\mathrm{Norm}_1(\mathbf{Z}^{(l-1)})\big),
\end{equation}
\begin{equation}
	\mathbf{Z}^{(l)} = \mathbf{H}^{(l)} + 
	\mathrm{FFN}\!\big(\mathrm{Norm}_2(\mathbf{H}^{(l)})\big),
\end{equation}
where $\mathrm{MHA}(\cdot)$ denotes multi-head self-attention, and $\mathrm{FFN}(\cdot)$ is the feed-forward module.


\paragraph{Multi-Head Self-Attention}
Given the normalized input $\mathbf{Z}\in\mathbb{R}^{ P\times D_{\mathrm{emb}}}$, the query, key, and value matrices are firstly computed as
\begin{equation}
	\mathbf{Q} = \mathbf{Z}\mathbf{W}_Q,\quad
	\mathbf{K} = \mathbf{Z}\mathbf{W}_K,\quad
	\mathbf{V} = \mathbf{Z}\mathbf{W}_V,
\end{equation}
where $\mathbf{W}_Q,\mathbf{W}_K,\mathbf{W}_V\in\mathbb{R}^{D_{\mathrm{emb}}\times D_{\mathrm{emb}}}$ are learnable projection matrices.

To realize multi-head attention, the embedding dimension is divided into $N_h$ heads, each of dimension $d_h=D_{\mathrm{emb}}/N_h$. Specifically, $\mathbf{Q},\mathbf{K},\mathbf{V}$ are reshaped into
\begin{equation}
	\mathbf{Q},\mathbf{K},\mathbf{V}\in\mathbb{R}^{ N_h\times P\times d_h},
\end{equation}
so that attention can be computed in parallel across all heads.

For the $j$-th head, the scaled dot-product attention is calculated as
\begin{equation}
	\mathrm{Attn}_j(\mathbf{Q}_j,\mathbf{K}_j,\mathbf{V}_j)
	= \mathrm{softmax}\!\left( \frac{\mathbf{Q}_j\mathbf{K}_j^\top}{\sqrt{d_h}} \right)\mathbf{V}_j,
\end{equation}
where $\mathbf{Q}_j,\mathbf{K}_j,\mathbf{V}_j \in \mathbb{R}^{P\times d_h}$ are the query, key, and value matrices corresponding to head $j$.

The outputs of all heads are then concatenated along the last dimension:
\begin{equation}
	\mathbf{A} = \mathrm{Concat}\big(\mathrm{Attn}_1,\mathrm{Attn}_2,\dots,\mathrm{Attn}_{N_h}\big) 
	\in \mathbb{R}^{ P\times D_{\mathrm{emb}}},
\end{equation}
and projected to the original embedding space through a linear layer (LL):
\begin{equation}
	 \mathrm{LL(}\mathbf{A})=\mathbf{AW}_O+\mathbf{b}_O,
\end{equation}
where $\mathbf{W}_O\in\mathbb{R}^{D_{\mathrm{emb}}\times D_{\mathrm{emb}}}$ is the learnable output projection matrix.

In summary, multi-head attention enables the model to jointly attend to information from multiple representation subspaces, enhancing its ability to capture diverse dependencies across the pilot sequence.

\paragraph{Feed-Forward Network and Normalization}
The FFN consists of two layers with ReLU activation in between:
\begin{equation}
	\mathrm{FFN}(\mathbf{x}) = \mathrm{ReLU}(\mathbf{x}\mathbf{W}_1+\mathbf{b}_1)\mathbf{W}_2+\mathbf{b}_2.
\end{equation}
where $\mathbf{x}$ denotes the input feature vector at a given time step, and $\mathbf{W}_1$, $\mathbf{W}_2$, $\mathbf{b}_1$, and $\mathbf{b}_2$ are trainable parameters.
Layer normalization is applied before both the attention and FFN modules, while residual connections ensure stable gradient propagation.

\paragraph{Output Layer.}
After passing through $L$ Transformer blocks, the sequence is normalized and projected to the codebook size. Specifically, for the last hidden representation $\mathbf{Z}^{(L)}$, the prediction logits are given by
\begin{equation}
	\mathbf{o}_p = \mathbf{z}^{(L)}_p \mathbf{W}_{\mathrm{out}} + \mathbf{b}_{\mathrm{out}},
	\quad \mathbf{o}_p \in \mathbb{R}^{|\mathcal{N}|},
\end{equation}
where $|\mathcal{N}|$ is the size of the near-field codebook. The predicted codeword index is obtained by applying a softmax classifier over $\mathbf{o}_p$. 

Thus, the GPT-2 module serves as the temporal predictor that maps a sequence of pilot-derived features into the most likely near-field beam index at future time instants.

\subsubsection{Model Training Strategy}
Inspired by the BERT paradigm in natural language processing, we adopt a two-stage training strategy consisting of pre-training and fine-tuning. 
The pre-training stage allows the model to learn general temporal–spatial representations from large-scale pilot data through a masking task, while the fine-tuning stage specializes the model for the downstream beam prediction objective.

\paragraph{Pre-training}
The objective of the pre-training stage is to expose the model to large amounts of pilot–beam correspondence data and to encourage it to learn temporal dependencies across multiple frames. Given the input pilot tensor 
\[
\mathbf{X}\in \mathbb{R}^{B\times P\times 2\times K\times G},
\]
where $B$ is the batch size. We first perform random masking over the temporal dimension. Specifically, let $\alpha$ denote the overall masking ratio. A proportion of $\alpha P$ frames are randomly selected as masked positions. Among these masked positions, 80\% are replaced by all-zero inputs, 10\% are substituted with randomly sampled pilot features, and the remaining 10\% are left unchanged. This design prevents the model from overfitting to a trivial ``mask token'' and encourages it to robustly capture both local and global temporal correlations.

After masking, the data is passed through the CNN feature extractor:
\[
\mathbf{F} = \mathrm{CNN}(\mathbf{X}^{\mathrm{masked}}) \in \mathbb{R}^{B\times P\times D_{\mathrm{out}}},
\]
and then into the GPT-2 module, which outputs the prediction logits:
\[
\mathbf{O} = \mathrm{GPT-2}(\mathbf{F}) \in \mathbb{R}^{B\times P\times |\mathcal{N}|},
\]
where $\mathbf{O}\in \mathbb{R}^{B\times P\times |\mathcal{N}|}$ stores the model’s unnormalized prediction scores, with indices $(b,p,n)$ specifying the batch index, temporal index, and codeword index, respectively. The training objective is the cross-entropy loss computed over all time steps:
\begin{equation}
\mathcal{L} _{\mathrm{pre}}=-\frac{1}{BP}\sum_{b=1}^B{\sum_{p=1}^P{\log \frac{\exp \bigl( \left[ \mathbf{O} \right] _{b,p,y_{b,p}} \bigr)}{\sum_{n=1}^{|\mathcal{N} |}{\exp \bigl( \left[ \mathbf{O} \right] _{b,p,n} \bigr)}},}},
\end{equation}
where $y_{b,p}$ denotes the ground-truth index of the optimal near-field codeword for user $b$ at frame $p$ and $\left[ \mathbf{O} \right] _{b,p,n}$ denotes the predicted logit value (before softmax) corresponding to the $n$-th codeword in the near-field codebook $\mathcal{N}$ 
for batch $b$ at frame $p$. 


\paragraph{Fine-tuning}
In the fine-tuning stage, the model is adapted for the actual prediction task: forecasting the near-field beam index at the next frame given the past $P$ frames. To this end, we only use the output at the last context position $P$:
\[
\left[ \mathbf{O} \right] _{b,P} \in \mathbb{R}^{|\mathcal{N}|},
\]
which already represents the prediction logits over the near-field codebook for batch $b$. The predicted codeword index for the $(P+1)$-th frame is then obtained as
\begin{equation}
\hat{y}_{b,P+1}=\mathrm{arg}\max_n \,\bigl( {\left[ \mathbf{O} \right] _{b,P,n}} \bigr) .
\end{equation}
The fine-tuning loss is defined as the cross-entropy between the predicted logits and the ground-truth index $y_{b,P+1}$:
\begin{equation}
\mathcal{L} _{\mathrm{fine}}=-\frac{1}{B}\sum_{b=1}^B{\log \frac{\exp \bigl( {\left[ \mathbf{O} \right] _{b,P,y_{b,P+1}}} \bigr)}{\sum_{n=1}^{|\mathcal{N} |}{\exp \bigl( {\left[ \mathbf{O} \right] _{b,P,n}} \bigr)}}.}
\end{equation}


In this way, pre-training equips the model with the ability to learn general pilot–beam patterns under noisy and incomplete observations, while fine-tuning specializes the model for next-frame beam prediction with improved accuracy and reduced overhead.

\paragraph{Advantages of the Training Strategy}
The proposed pre-training and fine-tuning strategy offers the following advantages:
\begin{itemize}
	\item \textbf{Dense supervision:} During pre-training, the model receives feedback from multiple temporal positions through the masking strategy, allowing it to capture richer sequential dependencies and converge more smoothly over time.
	\item \textbf{Robustness:} By adopting a denoising-style masking scheme (80/10/10), the network becomes less sensitive to missing or corrupted pilot signals, showing stronger resistance to noise and improved generalization across different channel conditions.
	\item \textbf{Label efficiency:} When fine-tuning, the pre-trained representation enables the model to adapt to new tasks with only a small number of labeled samples, greatly reducing data requirements and facilitating practical deployment in real systems.
\end{itemize}

\paragraph{Computational Complexity Analysis}
The computational complexity of the proposed CNN--GPT-2 framework consists of two main parts: the convolutional feature extraction and the GPT-2-based temporal modeling. For the CNN module, each convolutional layer involves operations with a complexity proportional to the product of the number of input channels $C_{\mathrm{in}}$, output channels $C_{\mathrm{out}}$, the spatial size of the feature map $H \times W$, and the squared kernel size ${K_{\mathrm{c}}}^2$. The complexity for a single layer is thus $\mathcal{O}\!\left(C_{\mathrm{in}} C_{\mathrm{out}}{K_{\mathrm{c}}}^2 H W \right)$. Since the number of convolutional layers $L_{\mathrm{CNN}}$ is fixed and small, and the kernel size ${K_{\mathrm{c}}}$ is typically small (e.g., 3), this component scales linearly with the input spatial size and the number of channels.

For the GPT-2 module, the dominant cost arises from the multi-head self-attention mechanism. Given a context length $P$, embedding dimension $D_{\mathrm{emb}}$, and number of layers $L_{\mathrm{GPT-2}}$, the complexity of each attention block is $\mathcal{O}(P^2 D_{\mathrm{emb}})$ due to the pairwise attention computation, while the feed-forward network within each layer adds an additional $\mathcal{O}(P D_{\mathrm{emb}}^2)$. Stacking $L_{\mathrm{GPT-2}}$ such blocks yields an overall complexity of $
\mathcal{O}\!\left(L_{\mathrm{GPT-2}} \left(P^2 D_{\mathrm{emb}} + P D_{\mathrm{emb}}^2 \right)\right).
$

Combining both modules, the overall computational complexity of the proposed framework is expressed as:
\begin{align}
	\mathcal{O}\!\bigg(\sum_{\ell=1}^{L_{\mathrm{CNN}}} &C_{\mathrm{in}}^{(\ell)} C_{\mathrm{out}}^{(\ell)} {K_{\mathrm{c}}}^{(\ell)2} H^{(\ell)} W^{(\ell)}\nonumber\\
	&+ L_{\mathrm{GPT-2}} \big(P^2 D_{\mathrm{emb}} + P D_{\mathrm{emb}}^2\big)\bigg).
\end{align}

\vspace{-0.5cm}

\section{Simulation Results}\label{simulation}

\begin{table}[t] 
	\centering
	\caption{Default system parameters.}
	\label{tab:channel_params}
	\resizebox{1\columnwidth}{!}{
		\begin{tabular}{p{4.5cm} p{5cm}}
			\hline
			\textbf{Parameter} & \textbf{Value} \\
			\hline
			Carrier frequency $f_c$ & $30\,\text{GHz}$ \\
			Subcarrier spacing & $120\,\text{kHz}$ \\
			Number of subcarriers $K$ & $60$ \\
			BS antenna elements $N_{\mathrm{BS}}$ & $256$ \\
			Antenna spacing $d$ & $0.5\lambda$ \\
			User number $U$ & $8$ \\
			Number of radio frames $P$ & $7$ \\
			Channel scenario & UMa LOS (TR~38.901) \\
			Uplink transmit power & $20\,\text{dBm}$ \\
			RF chains $N_{\mathrm{RF}}$ & $8$ \\
			Widebeam codebook size & $G=64$ \\
			Near-field codebook size & $|\mathcal{N}|=1280$ \\
			Number of clusters & $12$ (UMa LOS default) \\
			Rays per cluster & $20$ \\
			\hline
	\end{tabular}}
\vspace{-0.3cm}
\end{table}
In this section, we present simulation results to evaluate the performance of the proposed CNN–GPT-2 based near-field beam prediction scheme. The simulations are conducted under a standardized 5G NR frame structure and a channel model consistent with 3GPP TR 38.901 Release~17. We first describe the channel generation settings and network model configurations used in the experiments, and then provide detailed performance comparisons with conventional beam training baselines.
	\vspace{-0.4cm}

\subsection{Simulation Parameter Settings}
The simulation setup follows the 3GPP TR~38.901 Urban Macrocell (UMa) LOS scenario. 
Users are assumed to be randomly distributed within an angular range of $[-60^\circ, 60^\circ]$ 
and a distance range of $[5,20]$ meters from the base station, with velocities uniformly sampled 
from $[30,100]\,\mathrm{km/h}$. This ensures that both near-field propagation effects 
and mobility-induced channel variations are captured in the evaluation. The detailed channel parameters are summarized in Table~\ref{tab:channel_params}.

For pilot transmission, we adopt a widebeam analog precoding codebook, where each widebeam is obtained by combining $v=4$ adjacent DFT beams into a single probing beam. With $N_{\mathrm{BS}}=256$ antennas and $v=4$, the total widebeam codebook size is $G=64$. 
Hence, in each radio frame, only $G/N_{\mathrm{RF}}=64/8=8$ OFDM symbols are required to complete the widebeam sweeping process, which keeps the pilot overhead extremely low.
The near-field codebook is generated by jointly sampling the angular domain with 
$N_{\mathrm{BS}}=256$ antenna steering directions and the distance domain with five uniformly 
distributed sampling points in $[5,20]m$. 
Hence, the total size of the near-field codebook is $|\mathcal{N}|=1280$.

During training, the proposed CNN–GPT-2 framework first undergoes pre-training on $80$k samples, 
followed by fine-tuning on an additional $20$k$ $ samples. 
In contrast, all baseline models are directly trained on $100$k$ $ labeled samples without pre-training. 
In the pre-training stage, the CNN encoder and the first Transformer block are frozen to preserve 
low-level spatial feature extraction, while the remaining GPT-2 layers are fully trainable to capture 
temporal dependencies. 
This configuration balances parameter efficiency and representation adaptability, 
facilitating fast convergence during fine-tuning.

On the learning side, the network parameters are selected to balance modeling capacity and computational efficiency. The CNN module extracts features of dimension $D_{\text{out}}=512$, which is matched to the embedding dimension of the Transformer to enable seamless integration. The GPT-2 model is configured with $L=4$ layers and $N_h=8$ attention heads, which provides sufficient representational power for capturing long-term dependencies without incurring excessive complexity. 

The detailed model parameters are summarized in Table~\ref{network_table}.

\begin{table*}[t]
	\centering
	\caption{\text {CNN–GPT-2 Model Parameters}}
	\label{network_table}
	\resizebox{1.9\columnwidth}{!}{
		\begin{tabular}{cccc}
			\hline
			\textbf{Module} & \textbf{Network Layer} & \textbf{Structures} & \textbf{Number of Parameters} \\ \hline
			
			\multirow{5}{*}{Feature extraction module (CNN)} 
			& Initial Conv & $C_{\mathrm{in}}=2$, $C_{\mathrm{out}}=32$, $k=3\times 3$, ReLU, BatchNorm & $0.6\times 10^{4}$ \\
			& Residual Block 1 & $C_{\mathrm{in}}=32$, $C_{\mathrm{out}}=64$, stride=2, Dropout(0.2) & $7.4\times 10^{4}$ \\
			& Residual Block 2 & $C_{\mathrm{in}}=64$, $C_{\mathrm{out}}=64$, stride=1, Dropout(0.2) & $5.1\times 10^{4}$ \\
			& Residual Block 3 & $C_{\mathrm{in}}=64$, $C_{\mathrm{out}}=64$, stride=1, Dropout(0.2) & $5.1\times 10^{4}$ \\
			& Projection FC & Flatten $\rightarrow$ FC($1024 \to 512$), ReLU, Dropout(0.2) & $5.2\times 10^{5}$ \\
			\hline
			
			\multirow{3}{*}{Beam prediction module (GPT-2)} 
			& Embedding & Positional embedding: $P=7$, $D_{\mathrm{emb}}=512$ & $3.6\times 10^{3}$ \\
			& Transformer Blocks ($L=4$) & Each block: MHA ($N_h=8$, $d_h=64$), FFN ($512 \to 2048 \to 512$), LayerNorm, Dropout(0.2) & $\approx 2.5\times 10^{7}$ \\
			& Output Head & Linear($512 \to 1280$), Softmax & $6.5\times 10^{5}$ \\
			\hline
	\end{tabular}}
	\vspace{-0.3cm}
\end{table*}

	\vspace{-0.2cm}

\subsection{Evaluation Metrics and Baseline Algorithms}

To validate the effectiveness of the proposed CNN–GPT-2 based near-field beam prediction framework, we adopt three evaluation metrics and compare our approach with several representative baselines.
\paragraph{Evaluation Metrics}
\begin{itemize}
	\item \textbf{Prediction Accuracy:}  
	The ratio of correctly predicted beams over the total number of test samples. Formally, it is defined as
	\begin{equation}
		\mathrm{Acc} = \frac{1}{B}\sum_{b=1}^{B}\mathbf{1}\big( \hat{y}_b = y_b \big),
	\end{equation}
	where $B$ is the number of test samples, $y_b$ is the ground-truth optimal codeword index, $\hat{y}_b$ is the predicted index. Here, $\mathbf{1}(\cdot)$ denotes the indicator function, which outputs $1$ if the condition is satisfied and $0$ otherwise.


\item \textbf{Top-2 Accuracy:}  
This metric evaluates whether the true optimal codeword lies within the two most probable predictions. It is defined as
\begin{equation}
	\mathrm{Acc}_{\mathrm{Top2}} = \frac{1}{B}\sum_{b=1}^{B}\mathbf{1}\big( y_b \in \mathrm{Top2}(\hat{\mathbf{p}}_b) \big),
\end{equation}
where $\hat{\mathbf{p}}_b = [\mathbf{O}]_{b,P}$ denotes the predicted probability distribution over the codebook for sample $b$, 
and $\mathrm{Top2}(\cdot)$ returns the indices of the two highest-probability codewords.  
The Top-2 accuracy provides a practically meaningful measure: 
if the optimal beam is contained within the top two candidates, the BS can perform a lightweight second-stage beam training by probing only these two beams. 
This enables near-optimal beam alignment with minimal additional pilot overhead compared to exhaustive search.

	\item \textbf{Normalized Beamforming Gain (NBG):}  
	To assess link-level performance, we compute the normalized beamforming gain, which measures how close the predicted beam is to the optimal one in terms of received signal power:
	\begin{equation}
		\mathrm{NBG} = \frac{1}{K}\sum_{k=1}^{K}
		\frac{\left|\mathbf{h}_{u,k}^{\mathrm{dl}}\mathbf{f}_{u}^{\mathrm{pred}}\right|^2}
		{\max\limits_{\mathbf{f}_{u}^{\mathrm{RF}}\in\mathcal{N}}\left|\mathbf{h}_{u,k}^{\mathrm{dl}}\mathbf{f}_{u}^{\mathrm{RF}}\right|^2}.
	\end{equation}
	A higher NBG indicates that the predicted beam achieves signal power close to the optimal one on average across all subcarriers.
\end{itemize}


\begin{figure}[t]
	\centering
	\includegraphics[width=2.8in]{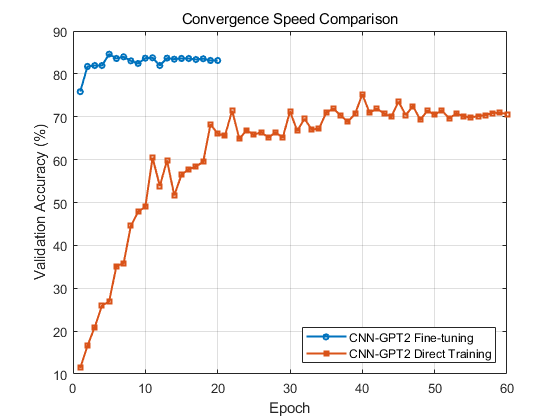}
	\vspace{-0.1cm}
	\caption{\fontsize{11pt}{\baselineskip}\selectfont Illustration of the convergence.}
	\label{epoch}
	\vspace{-0.5cm}
\end{figure}

\paragraph{Comparison Baselines}
We compare our proposed method with the following baselines:
\begin{itemize}
	\item \textbf{RNN-based beam prediction:} A recurrent neural network is trained to capture temporal dependencies in pilot sequences for predicting the optimal codeword index.
	\item \textbf{LSTM-based beam prediction:} A long short-term memory network is employed to improve long-term memory capability compared to vanilla RNNs.
	\item \textbf{GRU-based beam prediction:} A gated recurrent unit network is adopted as a lighter alternative to LSTMs, balancing performance and complexity.
\item \textbf{CNN–GPT-2 (direct training):}  
A variant of our proposed CNN–GPT-2 architecture that is trained directly on labeled beam prediction data without the pre-training–fine-tuning paradigm. This baseline highlights the benefits of the designed pre-training strategy by providing a fair comparison between direct supervised learning and the proposed two-stage training scheme.
\end{itemize}
\paragraph{Ablation Baselines}
To further analyze the effectiveness of each component in the proposed framework, we also consider the following ablation settings:

\begin{itemize}
	\item \textbf{Varying mask ratio $\alpha$:}  
	We investigate different masking ratios in the pre-training stage (e.g., $\alpha=0.2,0.4,0.6$) to evaluate how the amount of masked frames affects the representation learning and downstream beam prediction accuracy.
	
	\item \textbf{Supervision scope (Masked-only loss):}  
	Instead of computing the loss across all time steps, we consider the variant that only supervises the masked positions during pre-training. This ablation highlights the benefit of leveraging both masked and unmasked positions for stable optimization.
	
	\item \textbf{Simplified pilot transmission:}  
	We replace the proposed Zadoff–Chu based pilots and frequency-varying digital precoding with simplified pilots (all-ones sequence, identity digital precoder). This variant evaluates the performance gain brought by the carefully designed pilot transmission scheme.
%
\end{itemize}

\vspace{-0.3cm}

\subsection{Simulation Results}

In this subsection, we present the performance of the proposed CNN–GPT-2 based near-field beam prediction scheme and compare it with several baseline algorithms, including RNN, LSTM, GRU, and the direct training. Furthermore, we conduct ablation studies to evaluate the impact of key components such as the pilot transmission design and the pre-training strategy.


\begin{figure*}[t]
	\centering
	\subfigure[Prediction Accuracy versus Noise Power]{
		\label{acc}
		\includegraphics[width=0.32\linewidth]{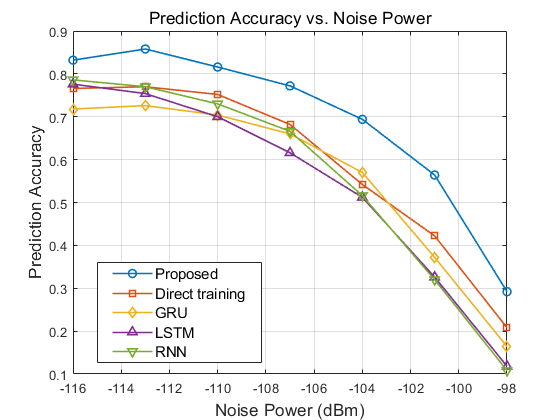}}
	\subfigure[Top-2 Accuracy versus Noise Power]{
		\label{top}
		\includegraphics[width=0.32\linewidth]{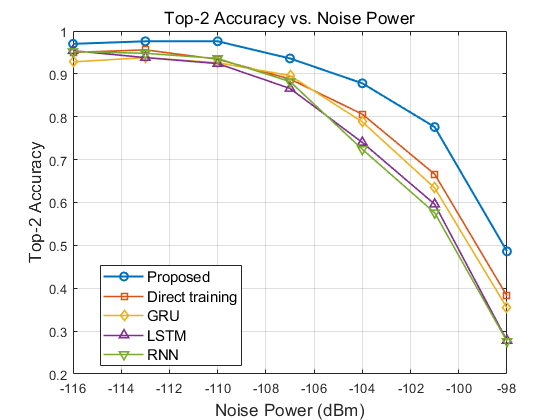}}
	\subfigure[NBG versus Noise Power]{
		\label{nbg}
		\includegraphics[width=0.32\linewidth]{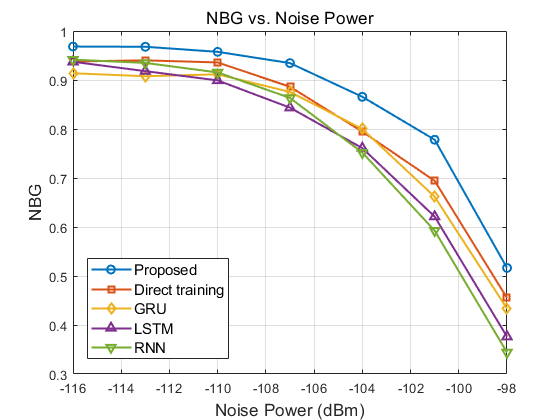}}
	\caption{Comparison of beam prediction performance among different baseline models under various noise power levels.}
	\label{fig:performance_noise}
	\vspace{-0.6cm}
\end{figure*}

Fig.~\ref{epoch} compares the convergence of the proposed pre-training + fine-tuning scheme with direct training. The proposed scheme rapidly reaches over $80\%$ validation accuracy within five epochs and stabilizes around $83\%\sim84\%$. In contrast, direct training requires more than 30 epochs to approach similar performance and remains unstable around $70\%$. These results confirm that pre-training provides a favorable initialization, leading to faster convergence, higher final accuracy, and reduced training cost.

The superior convergence of our approach highlights two main advantages. First, it reduces the overall training time and computational cost, which is crucial when handling large-scale near-field datasets. Second, it improves robustness to training instability, as evidenced by the stable accuracy curve compared to the oscillatory behavior of direct training.

Fig.~\ref{acc}  illustrates the prediction accuracy of different algorithms as a function of the uplink pilot noise power. The proposed CNN–GPT-2 with pre-training consistently achieves the highest accuracy across all noise levels, reaching over $85\%$ when the noise power decreases below $-113$ dBm. In contrast, conventional RNN, LSTM, and GRU models show slower performance growth with SNR, and their accuracy saturates below that of our scheme. The direct training CNN–GPT-2 baseline also lags behind, especially in the low-SNR regime, confirming that pre-training provides more robust initialization.

Fig.~\ref{top} presents the Top-2 prediction accuracy, which evaluates the probability that the true optimal beam is contained within the top two predicted candidates. This metric is particularly relevant in practice, as a small number of secondary pilot measurements can be used to identify the final optimal beam. The results show that the proposed CNN–GPT-2 scheme achieves near-perfect Top-2 accuracy (above $97\%$) when the noise power is below $-110$ dBm, substantially outperforming recurrent baselines. Even at higher noise levels, the proposed scheme maintains a clear advantage, reflecting its strong ability to capture temporal dependencies and generate a reliable candidate set for efficient near-field beam refinement.

Fig.~\ref{nbg} presents the normalized beamforming gain (NBG) performance under different uplink noise power values. The proposed CNN–GPT-2 model with pre-training achieves the highest NBG across all noise levels, exceeding $0.95$ when the noise power is below $-110$\,dBm. In contrast, the directly trained CNN–GPT-2 and recurrent baselines (RNN, LSTM, GRU) show a clear degradation, particularly in the low-SNR regime. The performance gap between the proposed and direct-training variants highlights the benefit of the pre-training stage in learning robust temporal dependencies from noisy pilot signals. Moreover, the relatively smooth NBG curve of the proposed model indicates stronger resilience to channel noise and better generalization across varying SNR conditions.

\begin{figure*}[t]
	\centering
	\subfigure[Prediction Accuracy versus Noise Power]{
		\label{acc_ab}
		\includegraphics[width=0.32\linewidth]{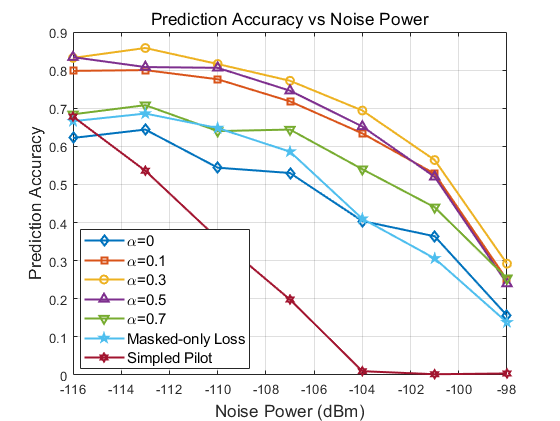}}
	\subfigure[Top-2 Accuracy versus Noise Power]{
		\label{top_ab}
		\includegraphics[width=0.32\linewidth]{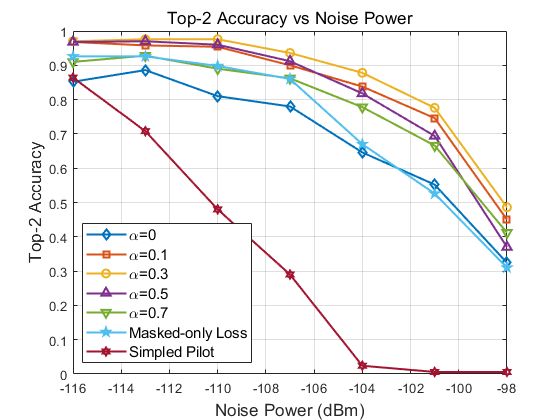}}
	\subfigure[NBG versus Noise Power]{
		\label{nbg_ab}
		\includegraphics[width=0.32\linewidth]{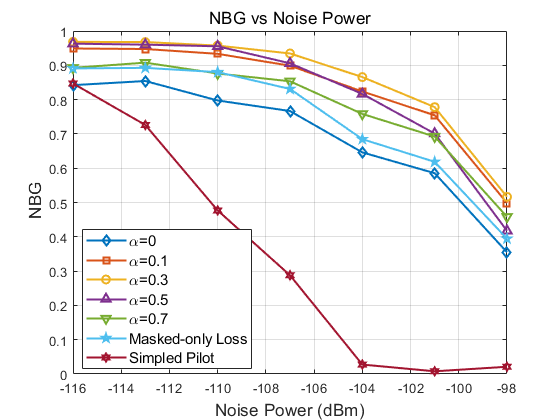}}
	\caption{Ablation study of the proposed CNN–GPT-2 framework under different configurations.}
	\label{fig:ablation_noise}
	\vspace{-0.6cm}
\end{figure*}

Fig.~\ref{acc_ab}, Fig.~\ref{top_ab}, and Fig.~\ref{nbg_ab} illustrate the ablation results under different masking ratios $\alpha$ and design variations.  
Specifically, Fig.~\ref{acc_ab} and Fig.~\ref{top_ab} show the prediction accuracy and top-2 accuracy versus noise power, while Fig.~\ref{nbg_ab} presents the corresponding normalized beamforming gain (NBG).  
The proposed configuration with $\alpha=0.3$ achieves the best overall performance across all noise levels, demonstrating that an appropriate masking ratio effectively balances information retention and contextual learning.  
When $\alpha$ is too small, the model receives insufficient self-supervised signal, limiting its generalization, whereas an excessively large $\alpha$ makes the prediction unstable due to the scarcity of visible context.  
The ``Masked-only Supervision’’ variant yields lower accuracy, confirming that incorporating both masked and unmasked frames in the loss computation stabilizes optimization and improves temporal consistency.  

Notably, the ``Simplified Pilot’’ configuration exhibits extremely poor performance at high noise levels (e.g., above $-102$\,dBm) but performs reasonably well in low-noise conditions.  
This behavior arises because, without Zadoff–Chu (ZC) sequences or frequency-varying precoding, the received pilots lack orthogonality and frequency diversity, making the model highly sensitive to noise and unable to distinguish effective channel features under low SNR.  
In contrast, the proposed pilot design maintains robustness and stable accuracy across all noise regimes, validating the effectiveness of the jointly optimized SRS-based pilot structure and end-to-end learning framework.


\begin{figure*}[t]
	\centering
	\subfigure[Prediction Accuracy versus User Speed]{
		\label{speed}
		\includegraphics[width=0.32\linewidth]{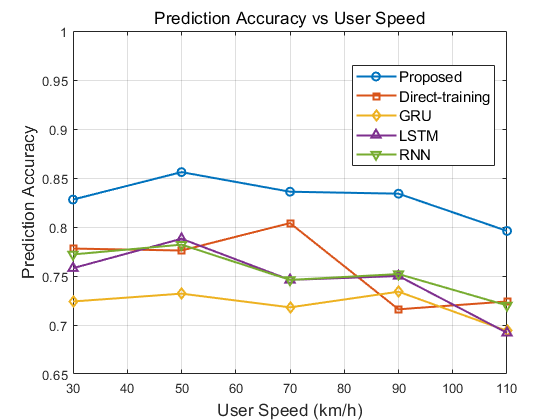}}
	\subfigure[Prediction Accuracy versus Channel $K_{R}$-Factor]{
		\label{acc_k}
		\includegraphics[width=0.32\linewidth]{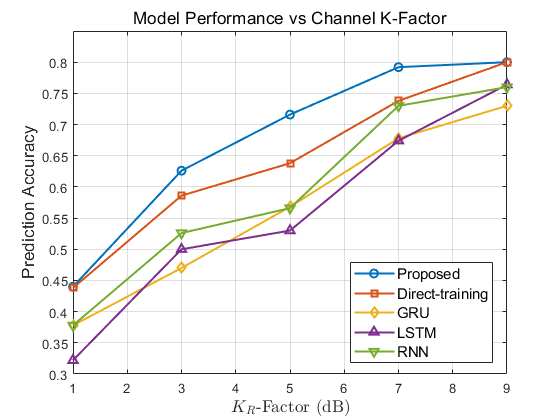}}
	\subfigure[Prediction Accuracy versus Noise Power under UMi Scenario]{
		\label{umi}
		\includegraphics[width=0.32\linewidth]{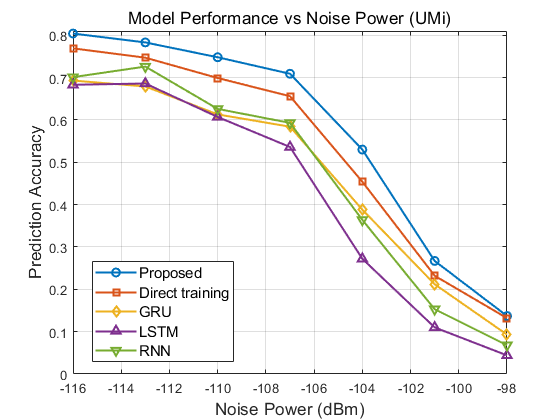}}
	\caption{Generalization evaluation of the proposed CNN–GPT-2 framework across different scenarios: (a) prediction performance under varying user speeds, (b) cross-channel adaptation under different $K$-factors, and (c) transferability to the UMi channel with varying noise power.}
	\label{fig:generalization}
	\vspace{-0.6cm}
\end{figure*}

Fig.~\ref{speed} illustrates the prediction accuracy of different models under varying user speeds.  
As the mobility increases from $30$\,km/h to $110$\,km/h, all schemes exhibit a gradual decline in performance due to the faster channel variations and reduced temporal correlation across frames.  
Nevertheless, the proposed CNN–GPT-2 model consistently outperforms the baseline methods at all speeds, maintaining an accuracy above $0.80$ even at $110$\,km/h.  
This robustness stems from the GPT-2 module’s strong temporal modeling capability, which effectively captures long-term dependencies and mitigates the channel decorrelation caused by user motion.  

\noindent\textbf{Model Generalization Across Channel Conditions.} 
Fig.~\ref{acc_k} illustrates the prediction accuracy of different schemes under channels with varying Rician $K_{R}$-factors. 
The proposed CNN–GPT-2 model was pre-trained on channels with $K_{R}=10$\,dB using $80$k samples, and subsequently fine-tuned with $20$k samples for each new $K_{R}$-factor, while all baseline models (CNN–GPT-2 with direct training, RNN, LSTM, and GRU) were trained from scratch with $100$k samples per condition. 
As shown in the figure, the proposed pretrain–finetune framework consistently achieves higher accuracy across a wide range of $K_{R}$-factors. 
In particular, the performance gain is more pronounced at lower $K_{R}$ values, where the channel exhibits stronger multipath components and higher nonlinearity. 
This demonstrates that the proposed model effectively captures the underlying temporal–spatial correlations of near-field propagation and generalizes well to unseen scattering conditions. 

\noindent\textbf{Cross-Scenario Transferability.} 
To further evaluate the cross-environment adaptability, Fig.~\ref{umi} compares the prediction accuracy under the 3GPP UMi channel model. 
The proposed model was first pre-trained on UMa channel data (80k samples) and then fine-tuned with only 20k UMi samples, whereas all baseline models were trained directly on 100k UMi samples. 
As observed, the proposed approach achieves consistently superior performance, particularly in the low-SNR regime, indicating its strong transferability across channel environments. 
This result highlights that the proposed large-model-based framework can effectively leverage knowledge learned from one propagation scenario to another, thereby exhibiting strong generalization and robustness to channel variations.

\vspace{-0.3cm}

\section{Conclusions}\label{conclusion}
This paper proposed an end-to-end CNN–GPT-2 framework for near-field beam prediction in extremely large-scale antenna systems. 
By integrating convolutional spatial feature extraction with GPT-2-based temporal modeling, the framework efficiently infers near-field beam indices from uplink pilots, achieving near-optimal performance without exhaustive beam search.
A BERT-inspired pre-training and fine-tuning strategy was developed to enhance convergence stability and generalization. 
Moreover, a practical uplink pilot transmission scheme combining Zadoff–Chu sequences, widebeam analog precoding, and frequency-dependent digital precoding effectively reduces pilot overhead while preserving rich spatial–frequency information. 
Simulation results under standardized 5G NR and 3GPP channel models confirmed that the proposed method achieves superior prediction accuracy and robustness compared with RNN-, LSTM-, and GRU-based baselines.


\bibliographystyle{IEEEtran}
\vspace{-0.4cm}
\bibliography{myre}

\end{document}